# Phonon Thermal Transport of URu$_2$Si$_2$: Broken Translational Symmetry and Strong-Coupling of the "Hidden Order" to the Lattice

P. A. Sharma[1], N. Harrison[1], M. Jaime[1], Y. S. Oh[2], K. H. Kim[2],
C. D. Batista[3], H. Amitsuka[4], and J. A. Mydosh[5,6,a)]

[1]*National High Magnetic Field Laboratory, Los Alamos National Laboratory, MS E536, Los Alamos, New Mexico 87545, USA*
[2]*CSCMR & School of Physics, Seoul National University, Seoul 151-742, Korea.*
[3]*Los Alamos National Laboratory, Theoretical Division, MS B262, Los Alamos, New Mexico 87545, USA*
[4]*Graduate School of Science, Hokkaido University, N10W8 Sapporo 060-0810, Japan*
[5]*Kamerlingh Onnes Laboratory, Leiden University, 2300RA Leiden, The Netherlands*
[6]*Max-Planck-Institut für Chemische Physik fester Stoffe, 01187 Dresden, Germany*

A dramatic increase in the total thermal conductivity ($\kappa$) is observed in the Hidden Order (HO) state of single crystal URu$_2$Si$_2$. Through measurements of the thermal Hall conductivity, we explicitly show that the electronic contribution to $\kappa$ is extremely small, so that this large increase in $\kappa$ is dominated by phonon conduction. An itinerant BCS/mean-field model describes this behavior well: the increase in $\kappa$ is associated with the opening of a large energy gap at the Fermi Surface, thereby decreasing electron-phonon scattering. Our analysis implies that the "Hidden Order" parameter is strongly coupled to the lattice, suggestive of a broken symmetry involving charge degrees of freedom.

At symmetry breaking (2$^{nd}$ order) phase transitions, an "order parameter" must characterize the non-symmetric, low temperature state [1]. The heavy-fermion URu$_2$Si$_2$ undergoes a 2$^{nd}$ order phase transition at ~ 17.5 K, yet the order parameter has remained "hidden" for over 20 years. This is an example of one of the most fundamental problems in solid state physics: the symmetries and associated order parameters are far from obvious in many interesting systems. In fact, the order parameters of antiferromagnets (the staggered magnetization) and superconductors (the phase of the condensate wavefunction) were similarly "hidden" for many years. New discoveries of such non-trivial hidden-order systems, such as spin glasses, or 2D XY magnets (where the "order" is of a purely topological character) thus continually expand our understanding of symmetry in solids. It for this reason that uncovering the "Hidden Order" (HO) phase in URu$_2$Si$_2$ is so interesting.

Almost all conceivable experimental probes have been utilized in an attempt to uncover the HO phase. Yet still, this diverse body of experimental data does not seem to convincingly converge to any existing model [2-4]. For example, this 2$^{nd}$ order transition seems to have some characteristics of antiferromagnetic (AF) ordering: a large peak appears in the specific heat. However, the observed small staggered-moment is inconsistent with the amount of entropy associated with the transition [4], and is instead likely due to a small AF impurity-phase [5]. In addition, due to strong hybridization between the conduction electrons and nearly-localized U-5f$^2$ moments, neither a purely local- nor itinerant-electron model can be applied unambiguously. In fact, there is strong evidence for a Fermi-liquid comprised of $\Gamma_5$ non-Kramers doublet degrees of freedom [6,7].

We present another route towards uncovering the HO in URu$_2$Si$_2$ via thermal conductivity ($\kappa$) measurements. Surprisingly, $\kappa(T)$ undergoes a very large increase at the HO transition. Although this feature has been observed some time ago [8], and then reported again recently [9], previous authors have not been able to model this effect in a way that elucidates the HO. By studying both pure and Rh-doped compounds, we have succeeded in explicitly showing that this feature is dominated by phonons. We apply a mean-field/BCS framework to describe $\kappa_{phonon}$ whereby electron-phonon (e-ph) scattering decreases as a gap ($\Delta$) in the electronic density of states (DOS) opens at the HO transition. An unusually large $\Delta(0)/k_B T_C$ ratio of ~4 is needed in order for the model to best describe the data, in agreement with STM estimates of $\Delta(T)$, that is suggestive of a coupling of the order parameter to the lattice. Our analysis suggests that an itinerant model associated with translational symmetry breaking of the charge degrees of freedom is needed in order to uncover the Hidden order parameter.

$\kappa(T)$ and the thermal Hall conductivity, $\kappa_{XY}$, are measured using the steady state method with two calibrated Cernox thermometers up to 15 T and down to 2 K with an absolute error of no more than 10 %. Single crystal U(Ru$_{1-x}$Rh$_x$)$_2$Si$_2$ samples are grown using the Czochralski method with two separate batches [10] yielding identical results. Rh is doped onto the Ru site at a concentration of up to 4% ($x$=0.04).

Fig. 1a shows $\kappa(T)$ for tetragonal U(Rh,Ru)$_2$Si$_2$. All samples are measured with the heat current ($Q$) //$c$. URu$_2$Si$_2$ displays a large upturn right at the HO transition. As the Rh concentration is increased, this feature becomes significantly reduced until it is entirely lost at 3% Rh doping, for which AF sets in at $T_N$. Unlike in the HO state, the AF state induced via Rh doping displays a tiny feature upon ordering. Increasing the doping to 4% destroys both HO and AF so that no feature is seen at all in $\kappa(T)$. For each sample, a transition is observed where it is expected according to the known phase diagram [10]. The absolute magnitude of $\kappa$ at $T>T_{HO}$ for each sample is very similar and within the error bars. This trend is unlike that of the electrical conductivity ($\sigma$), which noticeably increases with Rh doping. Judging from the behavior of the non-ordered 4% sample, it appears that $\kappa(T)$ sharply increases at the HO transition in $URu_2Si_2$. Fig. 1b shows $\kappa(T)$ for $Q//H//c$ and $Q//a$ in zero field for pure $URu_2Si_2$. Clearly, $\kappa_a(T)>\kappa_c(T)$, in contrast to $\sigma(T)$ for which $\sigma_c>\sigma_a$ [11]. Interestingly, as $H$ is applied, the peak in $\kappa(T)$ is reduced in a more gradual way than the effects of doping. $T_{HO}$ is also gradually reduced with $H//c$ in accordance with the $H$ vs. $T$ phase diagram [12].

Before investigating any implications of the observed effect for the HO, one must determine the dominant contribution(s) to $\kappa$. We argue here that the observed increase in $\kappa$ is dominated by phonons rather than electrons [13], suggested by the data shown in Fig. 1. As already mentioned, the $\kappa$ anisotropy and systematic change with Rh doping is opposite to that of $\sigma$. If electronic thermal conduction were important, the behavior of $\kappa$ should be more strongly correlated with that of $\sigma$. In addition, $\kappa$ obeys a $T^2$ power law behavior at low $T$ (inset Fig. 1a), which is consistent with dominant phonon conduction in the presence of e-ph scattering [15].

The thermal Hall (Righi-Leduc) conductivity ($\kappa_{XY}$) is the best way of ascertaining $\kappa_E$. Though the Weidemann-Franz law (WFL), $\kappa_E=L_0\sigma T$, is often used to estimate the electronic contribution, the results of Fig. 1 imply that $\kappa_{phonon}$ is dominant and so the WFL is highly uncertain. Assuming that e⁻-ph scattering dominates, the WFL fails because of the preponderance of small angle scattering events (phonon wavevector $q<<k_F$) at low, yet intermediate $T$, which disproportionately affect the electronic heat (as opposed to charge) current [14]. $\kappa_{XY}$ provides a direct measurement of electronic heat conduction since electrons moving perpendicular to $H$ will be deflected in a transverse direction given by the Lorentz force, whilst phonons will not [15, 16].

Our determination of $\kappa_E$ is shown in Fig. 2. In Fig. 2b, we have shown the raw $\Delta T_{Hall}$ signal, measured isothermally as a function of $H$ for $T$=2, 5, 10 and 20 K [17]. The thermal hall conductivity is calculated using the expression $\kappa_{XY} = (\nabla_{Hall}T/\nabla_a T)\kappa_c$ [16]. $\nabla_a T$ is the longitudinal $T$ gradient produced by a $Q(//a)$ identical to that which produces $\nabla_{Hall}T$. $\kappa_a$ (correspondingly $\nabla_a T$) is independent of $H//a$. $\kappa_c$ is assumed not to change with $H//a$ (in the low-$H$ limit) similar to what is observed in $M$, $\rho$, as well as $\kappa_a$. $\kappa_{XY}$ is shown in Fig. 2 (c), approximately increasing linearly from zero at low $H$.

The *longitudinal* $\kappa_E$ can be derived from the $\kappa_{XY}(H)$ plot using the assumption: $\kappa_{XY}/\sigma_{XY}=\kappa_{XX}/\sigma_{XX}=\kappa_e/\sigma_c$. This relationship is more robust than the WF law [16]. The quantity $\sigma_{XY}/\sigma_c=\rho_{XY}/\rho_a=\tan\Theta_H(T)$ is then measured (Fig. 2b, inset) to calculate $\kappa_E(H)$ at each $T$, which is extrapolated at low $H$ to 0 T to determine $\kappa_E$(0 T) (Fig. 2a). The error bars shown for the $\kappa_E(T)$ plot indicate the cumulative errors from each successive step of the analysis. At 2 and 5 K, the error bars are approximately the size of the symbols. Due to errors in geometry, the error in the absolute magnitude of $\kappa_E$ may be up to 50 %.

Clearly, $\kappa_E$ makes up a quite small proportion of the total $\kappa$, shown more clearly in the inset of Fig. 2a, along with $\kappa_E$ expected from the WFL. At high $T$, the WFL clearly overstates the magnitude given by our estimate of $\kappa_E$. However, the two estimates approach each other at low $T$ as expected. The WF $T$-dependence near $T_{HO}$ is also incorrect based on our results. $\kappa_E$ itself appears to also undergo an increase, though not precisely at $T_{HO}$ and of a much smaller magnitude than $\kappa_{phonon}$. This is consistent with the claim that the electronic mean free path increases at the transition [18]. The primary assumption in previous reports has been to attribute some significance to the Lorenz ratio $L=\kappa/\sigma T$ using the raw $\kappa$ data. Here we show that this contains a large phonon component that should not be meaningfully included in this quantity. In particular, we present the low-$H$ transverse Lorenz number $L_{XY}(T, \mu_0 H=2\text{T})=\kappa_{XY}/\sigma_{XY}$. The magnitude and $T$-dependence of $L_{XY}$ (and by implication, $L$) are very different from previous reports [8,9].

Since $\kappa_E$ plays no important role in the large $\kappa$ increase, we analyze our data assuming only phonon conduction. It is clear that some scattering process drastically changes at $T_{HO}$. There is independent evidence [19, 20] that a gap, $\Delta(T)$, opens over parts of the Fermi surface (FS) at the transition. This will naturally lead to a decrease in e-ph scattering that can be modeled using the BCS formalism [21]. In particular, the effect of $\Delta(T)$ on $\kappa(T)$ has been explored (BRT), and applied to superconductors (SC) [22].

In order to perform the analysis, we first fit $\kappa(T)$ [23] of the 4% Rh doped sample, for which no transition is observed. This sample serves as a "background" that yields an independent way to extract the magnitudes of the various scattering rates for the undoped sample. The modification to e-ph scattering via the BRT formalism is then applied to describe URu$_2$Si$_2$. Near $T_{HO}$, e-ph scattering appears to be the most important. The fit, along with the experimental data points are shown in Fig. 3a for the 4% sample.

For $T<T_{HO}$, $\Delta$ becomes non-zero, effectively removing carriers and inhibiting e-ph scattering. The BRT e-ph scattering rate is given by: $\tau_O^{-1} = g(\hbar\omega/k_B T, \Delta/k_B T)\times\tau_N^{-1}$; the scattering rate in the ordered state ($\tau_O^{-1}$) is equal to the normal state value ($\tau_N^{-1}$), renormalized by a function *g*. The function *g* accounts for the fact that phonons with energy $\hbar\omega < 2\Delta$ are no longer scattered by electrons, thereby reducing the scattering rate and causing an increase in $\kappa(T)$ [24]. This increase in $\kappa(T)$ then tracks $\Delta(T)$. Since we have fit all the other scattering rates to the 4% sample, the only parameter left to describe URu$_2$Si$_2$ is $\Delta(T)$.

An experimental STM estimate of $\Delta(T)$ (Fig. 3a inset) [19] is inserted into the BRT scattering rate to compute the expected $\kappa_{phonon}(T)$ (BRT/STM model). The results of the BRT/SRM model are shown in Fig. 3a and compared to the raw data. The increase at $T_{HO}$ is clearly present in the calculation, though the magnitude overestimates the experimental result. This is understandable because there is a significant residual DOS within the gap that continues to contribute to e-ph scattering, reducing $\kappa(T)$. In order to try to take into account a finite DOS for $E<\Delta$, we assume that Mattheisson's rule is approximately valid and add the experimental 4% $\kappa(T)$ in parallel with the calculation for the pure sample according to $\kappa_{TOTAL}^{-1} = \alpha\kappa_{4\%}^{-1} + (1-\alpha)\kappa_{BRT}^{-1}$, where $\alpha$ can be adjusted in varying amounts. The results for a few values of $\alpha$ are shown in Fig. 3a. The value of $\kappa$ is lowered towards the experimental magnitude, yet the *T*-dependence is still incorrect. However, as $\alpha$ is increased, the calculation comes closer to the experimental $\kappa(T)$ for the 2% sample shown in Fig. 1a. Namely, the peak in $\kappa$ becomes rapidly reduced with even small amounts of doping. This simulation implies that adding small amounts of Rh (naively adding electrons) strongly increases the DOS at certain places in the FS, which also must reduce $\Delta$ because $T_{HO}$ is sharply reduced. Note also that adding Rh increases $\sigma$ [10], implying a weakened coupling of the f- and conduction electrons. The minute amount of Rh needed to produce such a strong change in $\kappa$ may imply an unusual instability in the FS, which leads to the HO ground state.

We further explore this model by inserting the BCS formula for $\Delta(T)$ into the BRT scattering rate. As shown in Fig. 3b, the use of a BCS $\Delta(T)$ comes closer to the observed $\kappa$ increase (as well as the BRT/STM model) as the value of $\Delta(0)/k_B T_C$ is progressively increased from the "weak-coupling" limit (1.764) to a "strong-coupling" limit (e.g. 4). Furthermore, it is likely that $\Delta(0)/k_B T_C > 4$ because the residual DOS for $E < \Delta$ is ignored in the model. The value $\Delta(0)/k_B T_C \sim 4$ is considerably larger than typical "strong-coupled" superconductors ($\sim 2$ in Pb and $\sim 2.5$ in $K_3C_{60}$ [25]), and in particular, is beyond the scope of the Migdal-Eliahsberg theory [26]. However, typical CDW systems show $\Delta(0)/k_B T_C \sim 3\text{-}12$ [21, 27]. These very large values are usually interpreted as being a consequence of the enormous elastic energy costs associated with a coupling between the CDW and the lattice. Unlike the CDW ground state, $URu_2Si_2$ does not undergo a large lattice distortion at $T_{HO}$. However, the large value of $\Delta(0)/k_B T > 4$ (intermediate between a strong coupled SC and CDW) implies a similarly significant coupling to the lattice.

The fact that this mean field (BCS) model for $\kappa_{phonon}$ reproduces the general form of the experimental results is an important clue to the mechanism for the HO transition. In particular, this model is essentially of an itinerant nature, implying instabilities in the FS that open a gap. For example, both the SC and CDW ground states can be described using the BCS formalism [21]. In fact, a $\kappa_{phonon}$ increase at SC/CDW/SDW transitions has been observed to occur [14,28] in a similar way as observed here. One must therefore conclude that our results support an itinerant model of the HO. In combination with this conclusion, the large value of $\Delta(0)/k_B T_C$ ($\sim 4$) that is needed to describe our data strongly suggest that the Hidden order parameter must be strongly coupled to the lattice. This claim is consistent with anomalies detected in the elastic constants at $T_{HO}$, in particular the observed softening of a shear mode implies a lattice instability [29]. If the HO is associated with the modulation of charge degrees of freedom (e.g. U electrical quadrupolar moments implicit to $\Gamma_5$ doublets [6,7]) this would naturally lead to a strong coupling to the lattice due to Coulomb forces. Note that exotic d-wave models [2, 3] should not have such a strong coupling to the lattice because the associated charge distributions possess no net electrical moment. There may well be similarities between the HO and localized charge-modulated states, such as the Charge-Ordered (CO) state. Very similar increases in $\kappa(T)$ have been observed upon entering into various CO states in, e.g. the manganites, nickelates, and cuprates [30].

We wish to acknowledge Huang Ying Kai (FOM-ALMOS) for the sample preparation.

a) Present Address: II. Physikalisches Institut, Universitat zu Koln, Germany

K/mm at the same value of $Q$ and $T$. $H$ is reversed at each $T$ point in order to subtract off any unavoidable transverse gradient, thus measuring the $H$-antisymmetric $\nabla_{Hall}T$ signal. $\Delta T_{HALL}$ was linearly proportional to $Q$.

FIGURE CAPTIONS

**Fig. 1** (a) $\kappa(T)$ for U(Ru$_{1-x}$,Rh$_x$)$_2$Si$_2$. A dramatic increase in $\kappa$ is observed at the "hidden order" transition, sharply suppressed with Rh doping. Antiferromagnetic order is realized at $x=0.03$, with only a slight feature in $\kappa$. No ordering was observed at $x=0.04$. Inset: The pure sample displays a low-$T$ $T^2$ power law, indicating phonon conduction in the presence of e-ph scattering. (b) $\kappa(T)$ of pure URu$_2$Si$_2$ with heat ($Q$) applied along the $a$-axis (dashed line) and the $c$-axis (closed squares). The $H$-dependence of the $c$-axis $\kappa(T)$ is shown (closed symbols) up to 14 T, with $H//c$.

**Fig. 2** (a) The electronic contribution to $\kappa(T)$ calculated from the thermal Hall effect in pure URu$_2$Si$_2$ measured in $H$ up to 6 T (closed symbols). The zero field $\kappa_{\text{elec}}(T)$ was extrapolated from the high field data (open circles). At low $T$, the error bars are about the size of the symbols. Inset: Comparison of the total $\kappa_C(T)$, the electronic contribution calculated using the Weidemann-Franz law ($L_0\sigma_C T$), and the zero field data in the main panel ($\kappa_{elec}$). (b) Raw thermal Hall signal as a function of $H$ at fixed $T$. Inset: Electrical Hall angle as a function of $T$ at various fixed $H$. (c) Thermal Hall conductivity as a function of $H$ at various fixed $T$. Inset: Hall Lorenz ratio, $L_{XY}$, as a function of $T$ and normalized to $L_0$. For clarity, only selected datasets are shown for panels (b) and (c). Lines are guides for the eye.

**Fig. 3** (a) The BRT model (open circles, described in text) describes the observed increase in $\kappa(T)$ for URu$_2$Si$_2$ (solid line) very well. A Debye-Calloway approximation (dashed line) was fit to the experimental $\kappa(T)$ for the $x=0.04$ sample (open up-triangles).

The energy gap, $\Delta(T)$ as determined by Rodrigo *et al.* (inset) was directly inserted into the BRT model. $\kappa(T)$ for the $x$=0.04 sample was added in varying amounts ($\alpha$) in parallel to the BRT model, in order to simulate the effect of a finite DOS within the gap at low $T$. Interestingly, the $\alpha$=0.8 (solid down-triangles) curve resembles that of the 2% sample shown in Fig. 1, upper panel. (b) A mean field solution for $\Delta(T)$ was inserted into the BRT equation and compared to the experimental result and the BRT/STM model. The ratio $\Delta(0)/k_B T_C$ was adjusted in order to match the previous results. A "strong coupling" value of $\Delta(0)/k_B T_C \sim 4$ more closely describes the data.

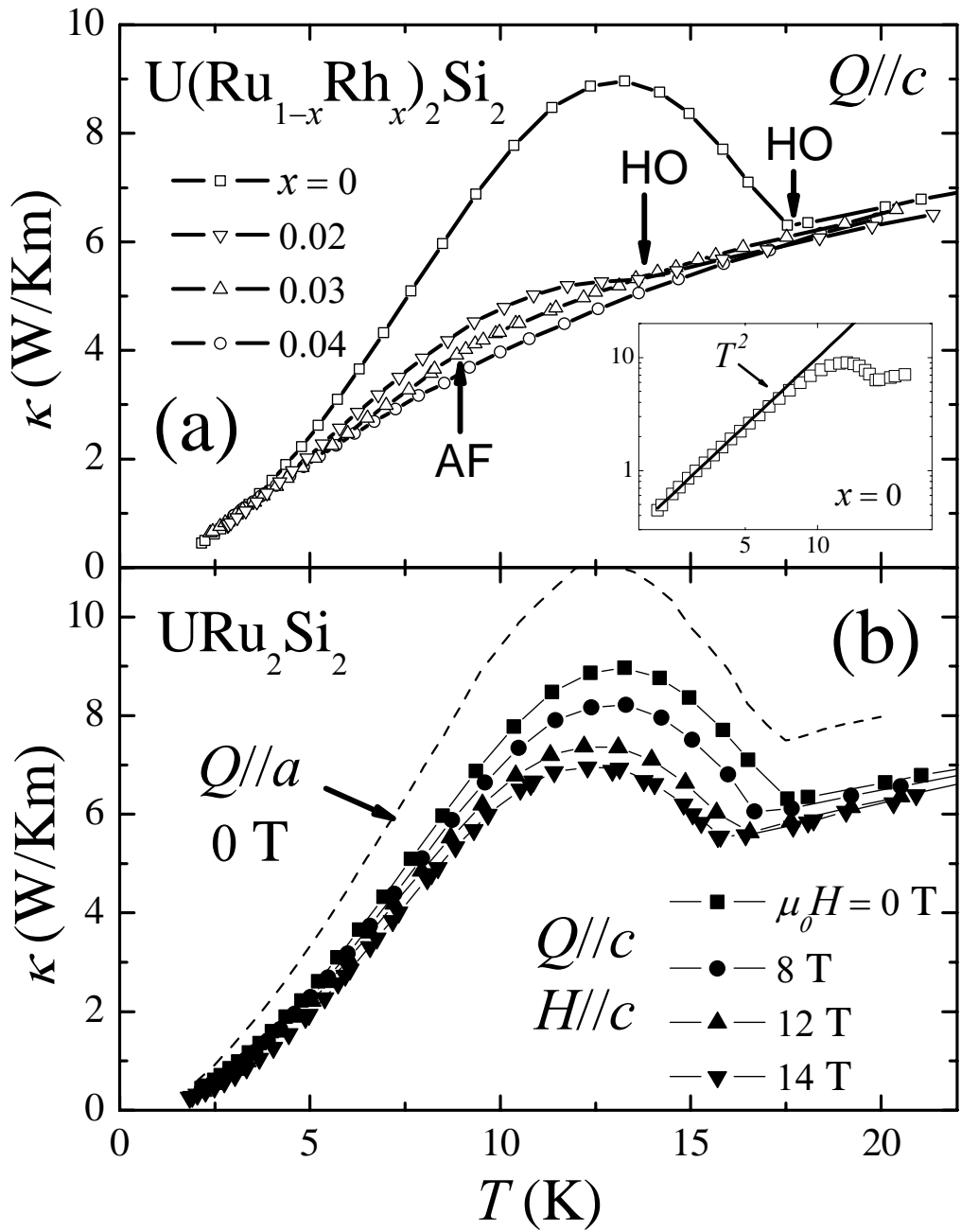

Figure 1, P. A. Sharma *et al.*

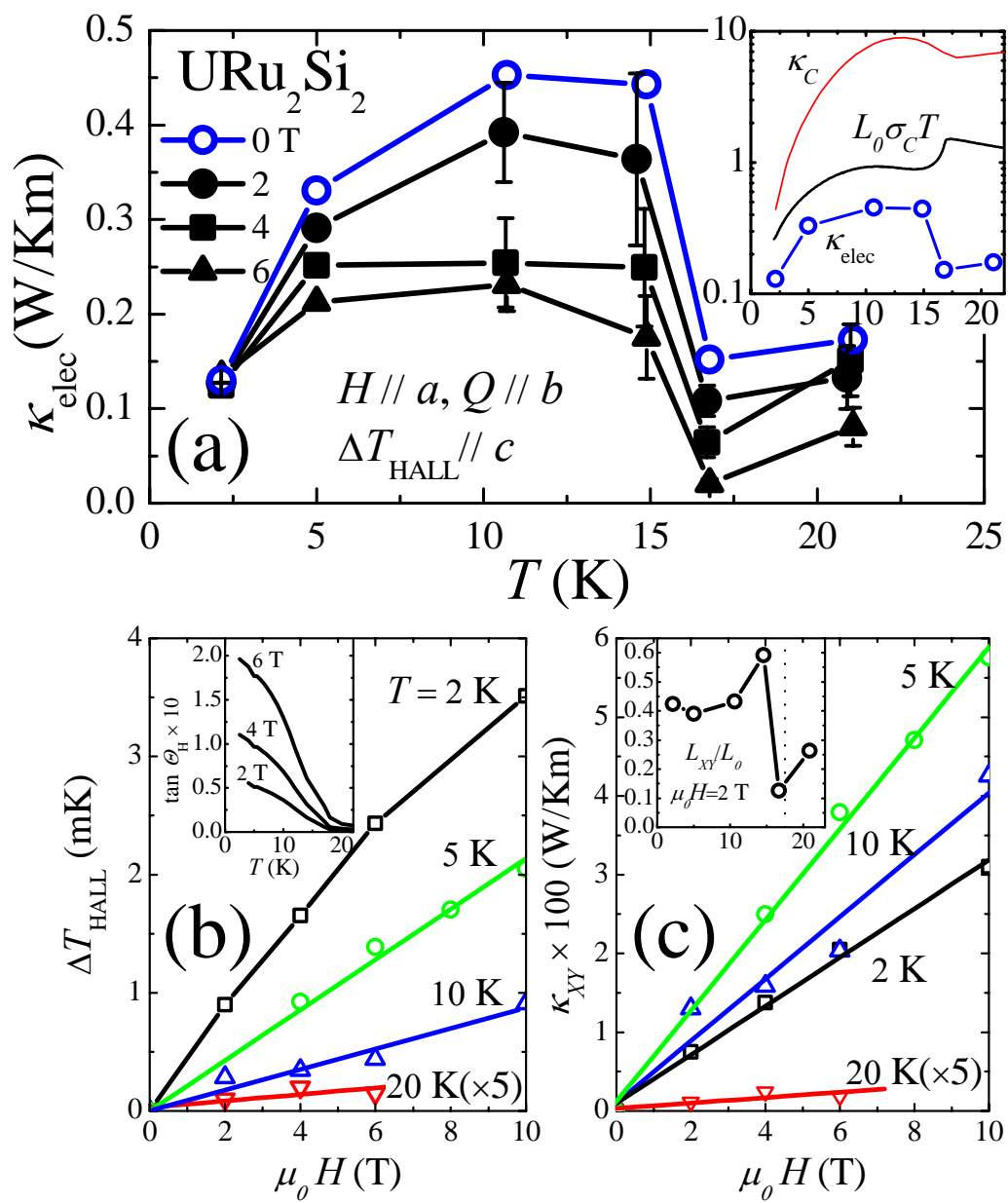

Figure 2, P. A. Sharma *et al.*

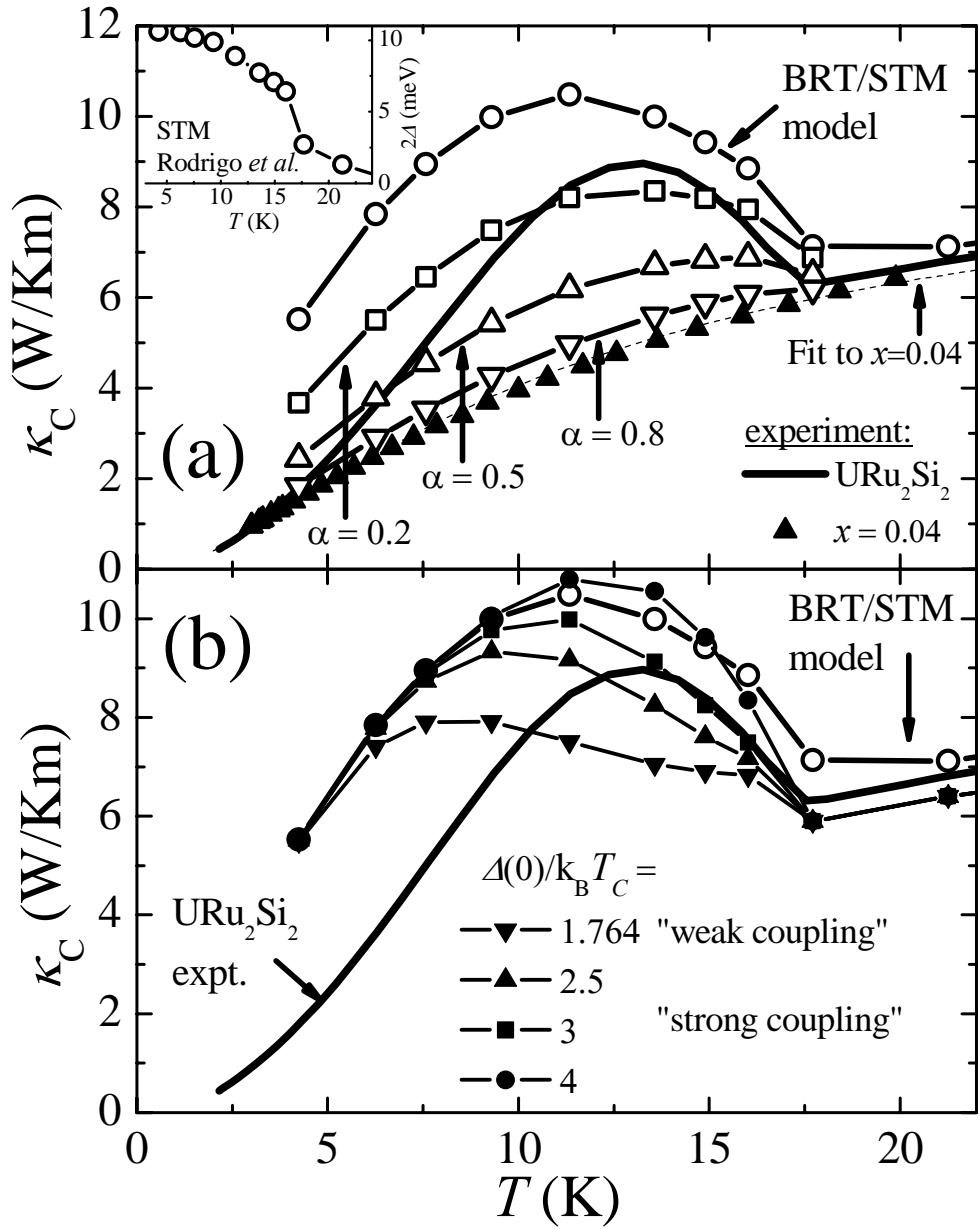

Figure 3, P. A. Sharma *et al.*